\documentclass[floatfix,aps,prb,showpacs,showkeys,groupedaddress]{revtex4}
\usepackage{graphicx}
\usepackage{amsmath}
\usepackage{amssymb}

\begin{document}

\title{Simulating the collapse transition of a two-dimensional semiflexible lattice polymer}

\author{Jie Zhou, Zhong-Can Ou-Yang, and Haijun Zhou}

\affiliation{Institute of Theoretical
  Physics, the Chinese Academy of Sciences,
Beijing 100080, China}


\begin{abstract}
  It has been revealed by mean-field theories and computer simulations that
  the nature of the collapse transition of a polymer is influenced
  by its bending stiffness $\epsilon_{\rm b}$. In two dimensions, a recent
  analytical work demonstrated that the collapse transition of a partially
  directed lattice polymer is always first-order as long as $\epsilon_{\rm b}$ is
  positive
  [H.~Zhou {\em et al.}, Phys.~Rev.~Lett.~{\bf 97}, 158302 (2006)].
  Here we employ Monte Carlo simulation to investigate systematically
  the effect of bending stiffness on the static properties of a 2D
  lattice polymer. The system's phase-diagram at zero force is obtained.
  Depending on $\epsilon_{\rm b}$ and the temperature $T$, the polymer can be
  in one of three phases: crystal, disordered globule, or swollen coil. The
  crystal-globule transition is discontinuous, the globule-coil transition is
  continuous. At moderate or high values of $\epsilon_{\rm b}$ the intermediate
  globular
  phase disappears and the polymer has only a discontinuous crystal-coil
  transition. When an external force is applied, the force-induced collapse
  transition will either be continuous or discontinuous, depending on whether
  the polymer is originally in the globular or the crystal phase at zero force.
  The simulation results also demonstrate an interesting scaling behavior of
  the polymer at the force-induced globule-coil transition.
\end{abstract}



\maketitle

\section{Introduction}
\label{sec:introduction}

A polymer is a linear chain of monomers
\cite{desCloizeaux-Jannink-1990}. The configuration of a long
polymer in solution is influenced by three different types of
interactions. Firstly, there are complicated interactions between
monomers such as hydrogen-bonding, weak van der Vaals attraction,
and electrostatic interactions. Monomer-monomer interactions can
also be mediated by solvent molecules, e.g., the attraction between
two negatively charged monomers due to the mediation of a
multi-valent positive charge (Mg$^{2+}$, for example). In
theoretical models of homologous polymers, the monomer-monomer
interactions can be represented by a potential energy $U({\bf r}_i,
{\bf r}_j)$ between two monomers at position ${\bf r}_i$ and
position ${\bf r}_j$. Monomer-monomer interactions can be either
attractive or repulsive. If the total interaction between a pair of
monomers is attractive, it tends to bring different parts of the
polymer together. At low temperatures, these monomer-monomer
attractive interactions result in the formation of compacted
(globular) polymer configurations. The volume of the polymer shrinks
as much as possible to minimize the total monomer-monomer contacting
energy. In such a globular configuration, the radius of gyration
$R_{\rm g}$ of the polymer scales with the polymer length $N$ as
$R_{\rm g}\propto N^{1/d}$, with $d$ being the
spatial dimension\cite{deGennes-1979}.
(For polymers in solution, $d=3$; for polymers
attached on the membrane of mobile lipids \cite{Maier-Raedler-1999},
$d=2$).

In the solution, solvent molecules collide frequently with the
polymer due to thermal agitation. These collisions tend to make the
polymer to take disordered and swollen coil configurations. At high
temperatures, thermal effects win over monomer-monomer attractions.
Consequently the polymer will take a randomly coiled shape, whose
radius of gyration scales with the polymer length $N$ as $R_{\rm g}
\propto N^{\nu}$, where the scaling exponent $\nu \approx 3/(d+2)$
according to the Flory mean-field theory
\cite{Flory-1967,deGennes-1979,Vanderzande-1998} and 2D exact
calculations \cite{Nienhuis-1982}.

Between the above-mentioned low-temperature globular phase and
high-temperature swollen coil phase, there exists also another
critical phase when the temperature is at a critical point
$T_\theta$, the so-called theta-point
\cite{desCloizeaux-Jannink-1990}. At $T_\theta$, the polymer
achieves a delicate balance among monomer-monomer attractions,
excluded-volume repulsions, and configurational entropy. The
polymer's gyration radius follow another new scaling law $R_{\rm g}
\propto N^{\nu_{\rm cr}}$, with the critical exponent $\nu_{\rm
cr}=4/7$ for 2D polymers
\cite{Duplantier-Saleur-1987,Vanderzande-1998} and $\nu_{\rm
cr}=1/2$ for 3D polymers \cite{Flory-1967,deGennes-1979}.

The third type of interactions are external forces. In recent
single-molecule manipulation experiments, external forces on the
order of piconewtons ($10^{-12}$ N) can be applied on the ends of a
polymer such as DNA \cite{Bustamante-etal-2003}. When the external
force field is sufficiently weak, the configuration of a
self-attracting polymer is not affected by the force, since the
attractive interaction between monomers is stronger to keep itself
remain compacted. On the other hand, when the external force is
sufficiently large, the polymer will be elongated considerably along
the force direction, and the end-to-end distance  of the polymer
chain scales {\em linearly} with polymer length $N$.

The overall shape of a self-attracting polymer can be dramatically
changed by changing environmental temperature and/or external
force. The globule-coil structural transition is a fundamental issue
in polymer physics. The nature of this transition has been studied
extensively in the past several decades by experimental,
theoretical, and simulation approaches
\cite{Zwanzig-Lauritzen-1968,Zwanzig-Lauritzen-1970,Massih-Moore-1975,Owczarek-etal-1994,Doniach-etal-1996,Doye-etal-1998,Kloczkowski-etal-1998,Owczarek-Prellberg-2000,Nowak-etal-2006,Kumar-etal-2007,deGennes-1979,desCloizeaux-Jannink-1990,Grosberg-Khokhlov-1994,Vanderzande-1998}.
Certain degree of consensus has been achieved on various aspects of
the polymer globule-coil transition, but there are still some
unresolved important issues. For a flexible self-attractive polymer,
mean-field theories predicted a second-order continuous globule-coil
phase-transition as the ambient temperature is elevated. This
prediction was confirmed by more recent analytical calculations
\cite{Marenduzzo-etal-2003,Rosa-etal-2006,Zhou-etal-2006} in 2D;
however, there are still some controversies in the literature
concerning Monte Carlo simulations in 3D (for example,
Refs.~[\onlinecite{Rampf-etal-2005,Rampf-etal-2006}] believed that the 3D
globule-coil transition is a first-order phase-transition in the
thermodynamic limit).  The force-induced globule-coil transition is
found to be first-order in 3D and to be second-order in 2D
\cite{Grassberger-Hsu-2002,Marenduzzo-etal-2003}.

The phase behavior of a semiflexible polymer is even more complex.
For a semiflexible polymer at very low temperature, the arrangement
of its segments in a globular conformation may be highly ordered.
This is driven by the desire of minimizing the total bending energy.
The polymer are therefore in a compact crystal phase. As temperature
increases, this crystalline order may be destroyed so as to gain
configurational entropy, but the polymer is still in a (disordered)
globular form. This crystal-globule transition is predicted to be a
first-order phase-transition both by exactly solvable models
\cite{Zwanzig-Lauritzen-1968,Zwanzig-Lauritzen-1970} and by $3D$
mean-field theory and Monte-Carlo simulation studies
\cite{Doniach-etal-1996,Doye-etal-1998,Lai-1998}. When the
temperature is further increased to the $\theta$ temperature, the
polymer globular conformation will change again to a swollen coil
conformation. This globule-coil transition is believed to be
second-order
\cite{deGennes-1979,Wittkop-etal-1995,Doniach-etal-1996,Doye-etal-1998,Rampf-etal-2005,Parsonsa-etal-2006}.
The $\theta$ temperature appears to be insensitive to the polymer's
bending stiffness
\cite{Doniach-etal-1996,Bastolla-Grassberger-1997,Doye-etal-1998}.
However, if the bending stiffness of the polymer is sufficiently
large, the solid-globule and globule-coil transitions may occur at
the same temperature, resulting in a single first-order solid-coil
structural transition \cite{Doniach-etal-1996}.

The force-induced collapse-transition have also been studied in
recent years. For 3D polymers, the transition is shown to be
first-order no matter whether the polymer is flexible or
semiflexible \cite{Grassberger-Hsu-2002,Rosa-etal-2006}. For 2D
flexible polymer, this transition is believed to be second-order
\cite{Grassberger-Hsu-2002,Marenduzzo-etal-2003,Marenduzzo-etal-2004};
while the results of an exactly solvable model \cite{Zhou-etal-2006}
suggest that the order will change to be first-order when the
polymer is semiflexible.

In this work, through a series of Monte Carlo simulations we
aim at a comprehensive understanding of the phase behavior of a 2D
semiflexible polymer under the action of temperature and external
force. We pay special attention on the effect of bending stiffness.
First we obtain the phase-diagram of the polymer at zero force using
temperature $T$ and bending stiffness $\epsilon_{\rm b}$ as  two
control parameters. At low $\epsilon_{\rm b}$, the polymer can
reside in three phases: crystal, disordered globule, and swollen
coil. The crystal-globule transition is first-order and the
globule-coil transition is second-order. When $\epsilon_{\rm b}$
exceeds some threshold value, however, the intermediate globular
phase of the polymer disappears. The polymer then has only one
crystal-coil transition, which is first-order. After the zero-force
phase-diagram of the polymer is known, we continue to study its
stretching behavior. We find that if at zero-force the polymer is in
the crystal phase, the applied force will cause a first-order
crystal-coil transition. But if the polymer is originally in the
disordered globule phase, the force-induced collapse transition will
be continuous. In the later case, the critical scaling behavior of
the extension of the polymer with chain length at the transition
point is obtained by MC simulation. To predict this critical scaling
behavior analytically requires further work.
With these results, the present simulation work enriches our
understanding of the structural properties of 2D self-attractive
polymers. It contributes to a full characterization of semiflexible
polymers both in 3D and on a surface.

The paper is organized as follows: The next section introduces the
model system; and the Monte Carlo simulation method is described in
Sec.~\ref{sec:MC-method}. Section~\ref{sec:temperature} and
Sec.~\ref{sec:force} focus, respectively, on temperature- and
force-induced structural transitions. We conclude this work in
Sec.~\ref{sec:conclusion}.


\section{The Model}
\label{sec:model}

\begin{figure} 
 \includegraphics[angle=0,width=0.5\textwidth]{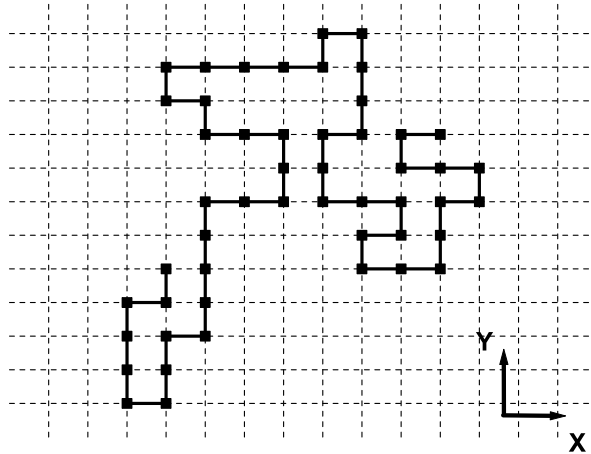}
  \caption{\label{fig:lattice_polymer}
    A polymer on a 2D square lattice with 
    lattice constant $a$. The configuration
    of the polymer fluctuates over time under 
    the joint action of
    monomer-monomer contacting interactions, 
    bending energy, external force, and  thermal energy.
  }
\end{figure}

We study by MC simulations the phase behaviors of a 2D self-avoiding
lattice polymer. The polymer is on a square lattice with edge length
$a$. The bond length between two consecutive monomers $i$ and $i+1$
of the polymer also equals to $a$. One end of the polymer is fixed;
the other end is free (without external force) or is being stretched
by an external force $f$ along the $x$-axis of the surface (see
Fig.~\ref{fig:lattice_polymer}).

If two monomers $i$ and $j$ of the polymer (with $|j-i|\geq 2$)
occupy nearest-neighboring sites on the square lattice, there is an
attractive energy of magnitude $\epsilon$. If in a configuration
${\cal C}$ there are a total number of $N_c$ such contacts, then the
total contacting energy is $-N_c \epsilon$. For semiflexible
polymers, there are also bending energies. In the lattice polymer,
whenever the configuration ${\cal C}$ makes a turn (with three
consecutive monomers not staying on a rectilinear line) there is an
energy penalty $\epsilon_{\rm b}$. Denote the total number of turns
in the configuration ${\cal C}$ as $N_{\rm b}$, the total bending
energy is then $N_{\rm b} \epsilon_{\rm b}$. The energy contributed
by the external force is equal to $- f x$, where $x$ is the $x$-axis
component of the end-to-end distance vector of the polymer. For a
given configuration ${\cal C}$, the total configurational energy is
expressed as
\begin{equation}
  \label{EQ:hamiltonian01}
  E( {\cal C}) =- N_c \epsilon + N_{b} \varepsilon_{b} -f x \ .
\end{equation}
The partition function of the polymer is
\begin{equation}
  \label{EQ:hamiltonian02}
  Z=\sum\limits_{{\cal C}} \exp\Biggl( -\frac{ E({\cal C}) }{ k_{B} T  } \Biggr) \ ,
\end{equation}
where $k_B$ is the Boltzmann constant and $T$ is the environmental
temperature.

In the following MC simulations, energy unit is set to be
$\epsilon$, length unit is set to be $a$, $k_B$ is set to be unity,
and force unit is set to be $\epsilon/a$. In the absence of external
force, the 3D version of this lattice-polymer was previously studied
by mean-field theory and by MC simulations in
Ref.~[\onlinecite{Doniach-etal-1996}].


\section{Monte Carlo simulation method}
\label{sec:MC-method}

Many different Monte Carlo procedures are documented in the
literature (see, e.g.,
Refs.~[\onlinecite{Rosenbluth-etal-1955,Grassberger-1997,Dai-etal-2003,Liang-etal-2003,Rampf-etal-2006,Huang-2007,Luo-etal-2007}])
to sample configurations of a polymer according to the Boltzmann
distribution
\begin{equation}
  \label{eq:Boltzmann}
  p({\cal C}) = \frac{ \exp\bigl(- E({\cal C})/ (k_B T) \bigr) }{ Z}
\end{equation}
where $Z$ is the total partition function of the polymer as given by
Eq.~(\ref{EQ:hamiltonian02}). The present MC procedure, which is
based on the idea of importance sampling (see, e.g., textbook
[\onlinecite{Newman-Barkema-1999}]), is inspired mainly by the earlier work
of
Refs.~[\onlinecite{Rosenbluth-etal-1955,Grassberger-1997,Doye-etal-1998,Dai-etal-2003}].
In sampling polymer configurations, the transition probability
$T(\mu\to \nu)$ for the polymer to evolve from an old configuration
$\mu$ to a new configuration $\nu$ is given by
\cite{Newman-Barkema-1999}:
\begin{equation}
    T(\mu\to \nu)=g(\mu\to
    \nu)\times A(\mu\to \nu) \ .
\end{equation}
In this equation, $g(\mu\to \nu)$ is the probability of proposing a
particular updated configuration $\nu$ from the old configuration
$\mu$, and $A(\mu\to \nu)$ is the probability of  accepting this
updates. To ensure detailed balance, $A(\mu\to \nu)$ is chosen such
that
\begin{equation}
    A(\mu\to \nu)=  \left \{
    \begin{array}{ll}
         \eta(\mu\to \nu) e^{-\beta (E_{\nu}-E_{\mu})} \ ,  &
    \textrm{if \;  $\eta(\mu\to \nu) e^{-\beta(E_{\nu}- E_{\mu})} < 1$}  \\
        1 \ , & \textrm{if \; $\eta(\mu\to \nu) e^{-\beta(E_{\nu}- E_{\mu})} \geq 1$} \\
    \end{array}
        \right.
        \label{eq:mc-rule}
\end{equation}
where $\eta(\mu\to \nu)\equiv g(\nu\to \mu)/ g(\mu\to \nu)$.

\begin{figure}
\includegraphics[angle=0, width=0.3\textwidth]{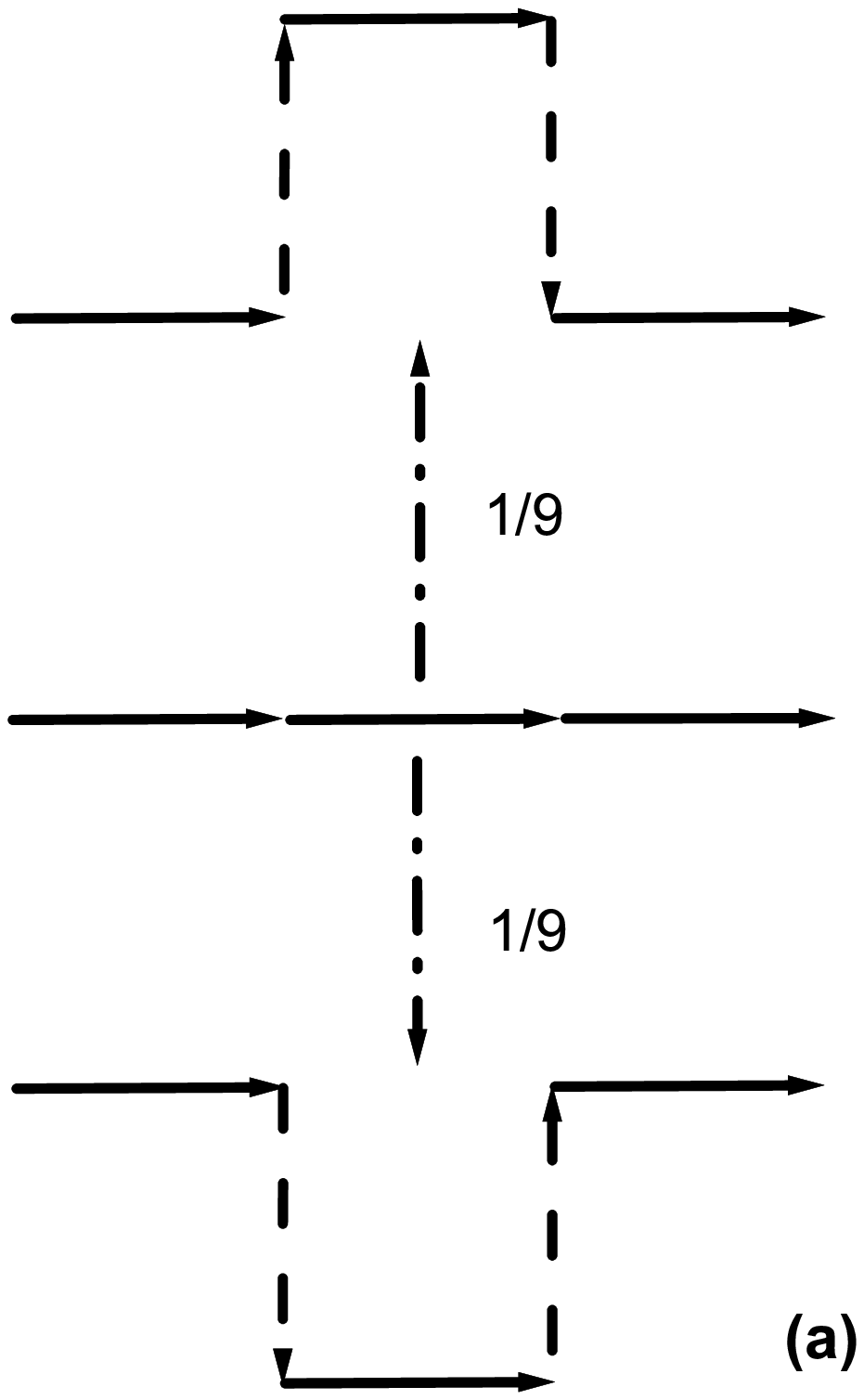}
\includegraphics[angle=0, width=0.3\textwidth]{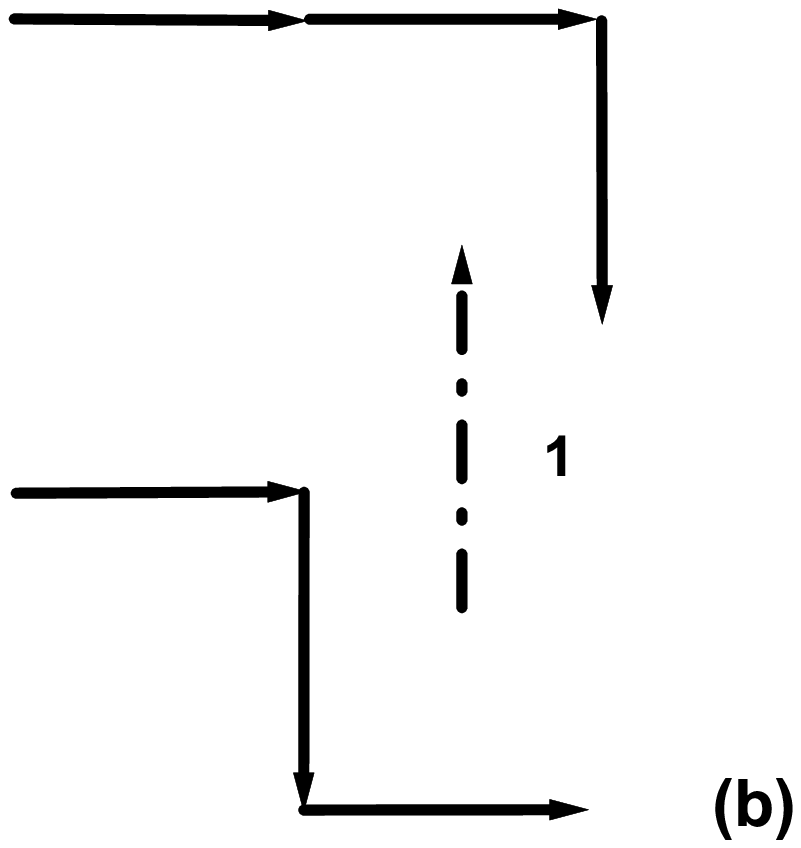}
\includegraphics[angle=0, width=0.3\textwidth]{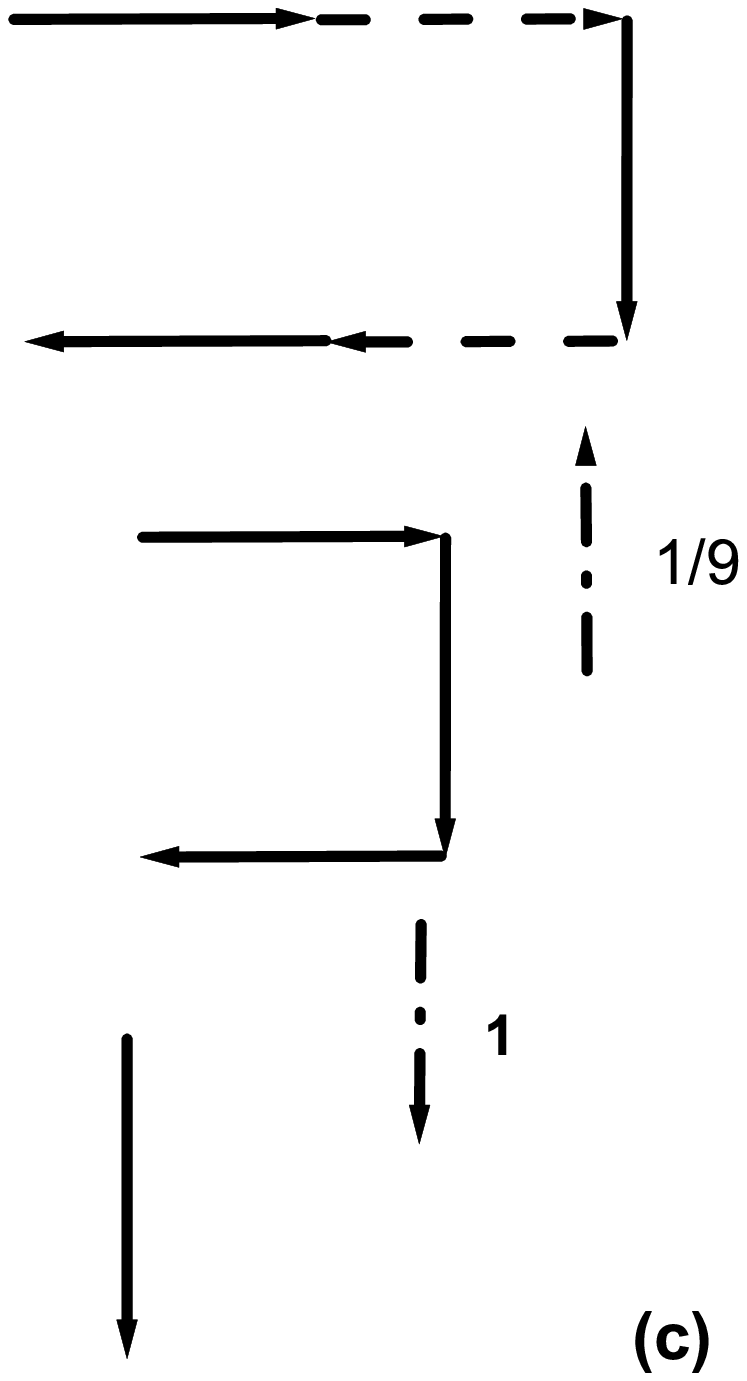}
\includegraphics[angle=0, width=0.3\textwidth]{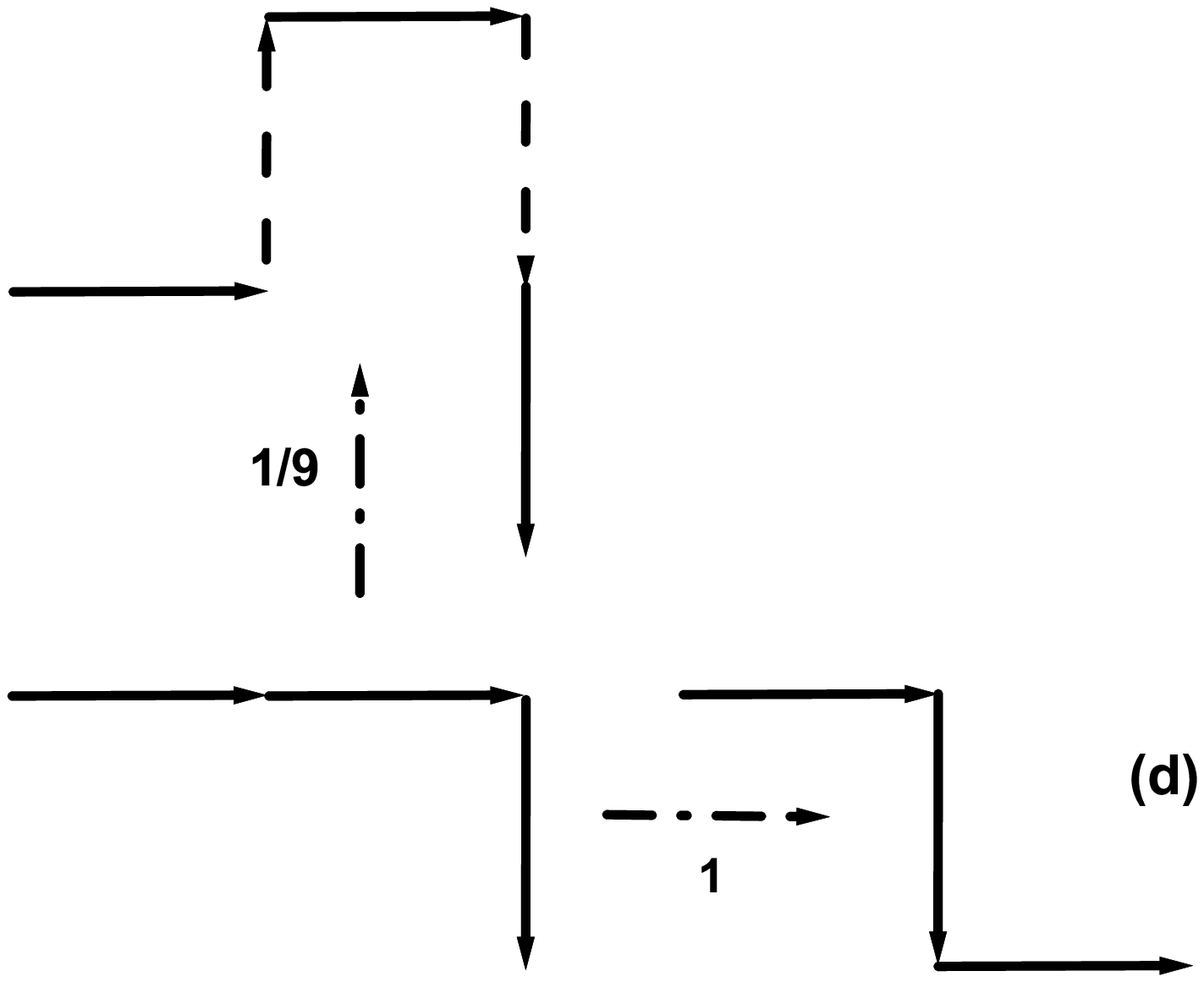}
\includegraphics[angle=0, width=0.3\textwidth]{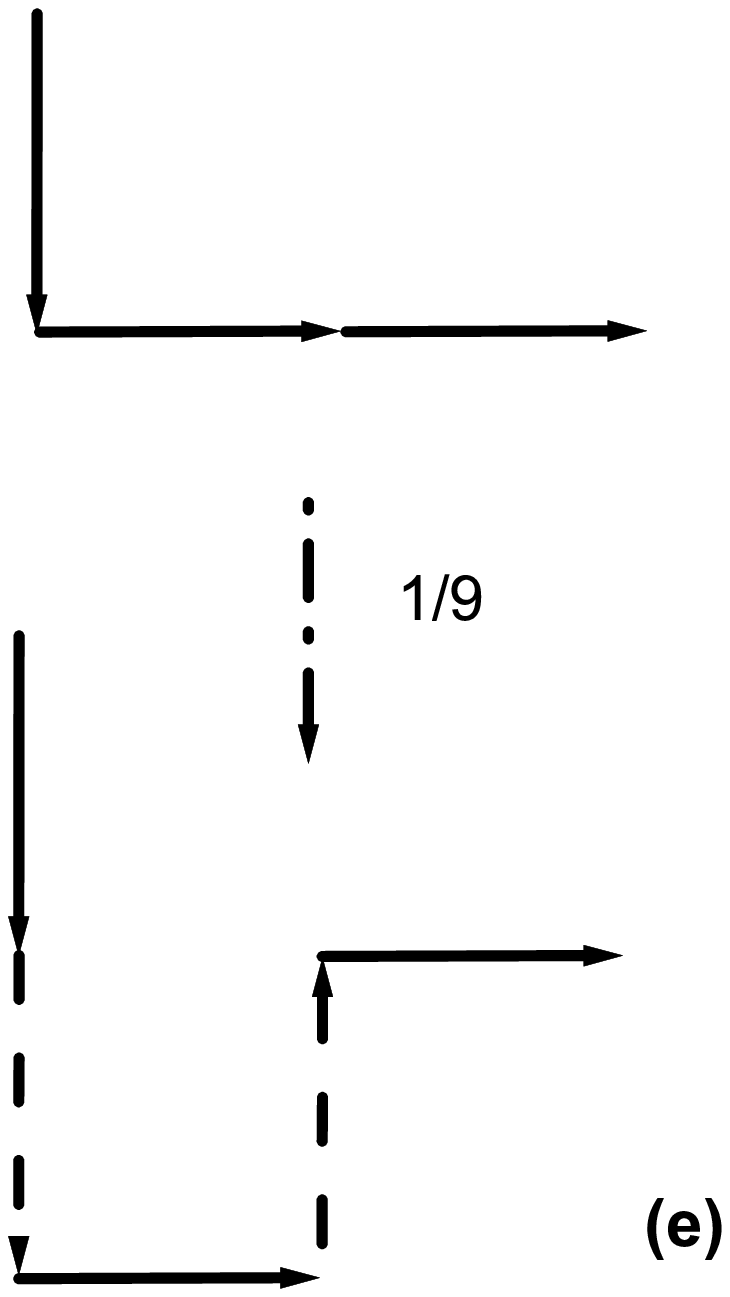}
\includegraphics[angle=0, width=0.3\textwidth]{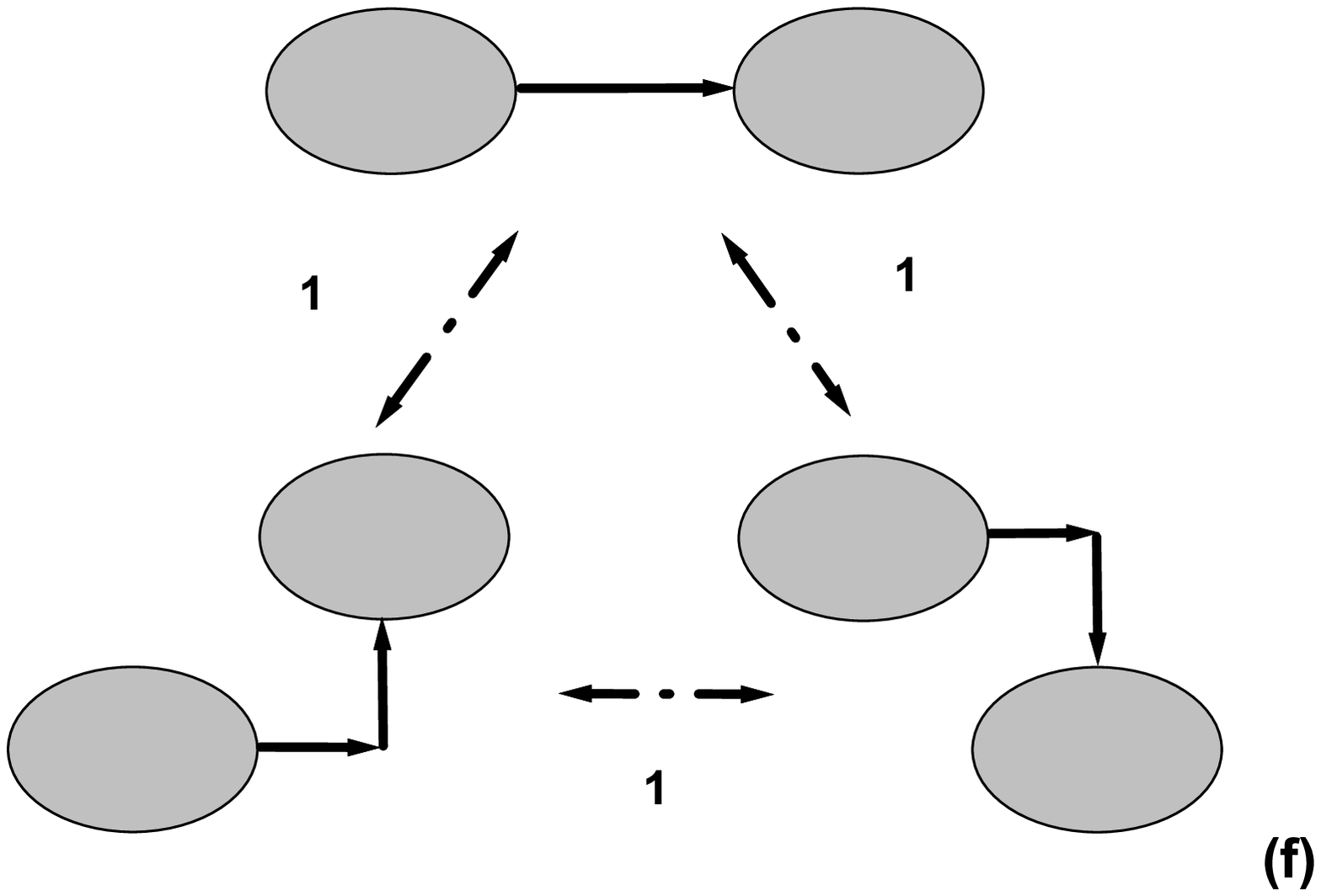}
\caption{\label{fig:operation} 
    Six different types of elementary configurational updates of the
    present MC simulation method.
  }
\end{figure}

In this paper, we use six different types of elementary updating
rules for the polymer's configuration (Fig.~\ref{fig:operation}).
These updating rules are described in some detail here.

\begin{enumerate}
\item[rule-1] Bonds $i-1$, $i$, and $i+1$ are parallel to each other
  [Fig.~\ref{fig:operation}(a)]. After the update,
  a hairpin is formed right after bond $i-1$, with
bond $i+1$ being at the head of the hairpin.  The polymer segment
from bond $i+3$ to bond $N$ reptates along its old contour and the
free end of the polymer shrinks by two bonds.

\item[rule-2] Bond $i$ is perpendicular to bonds $i-1$ and $i+1$
  [Fig.~\ref{fig:operation}(b)]. After the update, $i$ becomes parallel to bond $i-1$,
  while bond $i+1$ is perpendicular to both bond $i-1$ and bond $i$.

\item[rule-3] Bonds $i-1$, $i$, and $i+1$ forms a hairpin, with bond $i$ being
  at the head [Fig.~\ref{fig:operation}(c)]. After the update, with probability one-half,
  the stem length of this hairpin increases or decreases by one bond length.
  This stem length change is achieved by reptation, with the free end of the
  polymer draws back or stretches out by two bonds. If the free end of the polymer
  stretches out, the orientations of the last two bonds are randomly and
  independently assigned.

\item[rule-4] Bond $i+1$ is perpendicular to both bonds $i-1$ and $i$
  [Fig.~\ref{fig:operation}(d)].
  With probability one-half, after the update bond $i$ becomes perpendicular
  to both bonds $i-1$ and $i+1$. With the remaining probability one-half,
  after the update bond $i+1$ becomes the head of a hairpin while the polymer
  segment from bond $i+3$ to bond $N$ reptates along its old contour, and the
 free end of the polymer shrinks by two bonds.

\item[rule-5] Bond $i-1$ is perpendicular to both bonds $i$ and $i+1$
  [Fig.~\ref{fig:operation}(e)]. After the update, bond $i+1$
  becomes the head of a hairpin, while the polymer segment from bond $i+3$ to
$N$ reptates along its old contour and the free end of the polymer
shrinks by two bonds.

\item[rule-6] The polymer segment from bond $i+1$ to bond $N$ is rotated as a whole
with respect to bond $i$ clockwise or counter-clockwise by an angle
of $\pi /2$ or $\pi$ [Fig.~\ref{fig:operation}(f)]. This elementary
update causes a large change in the chain's configuration.
\end{enumerate}

We have taken special care to ensure that during these MC updating
processes, (1) no two monomers occupy the same site, (2) detailed
balance is not violated. To ensure detailed balance, the ratios
$\eta(\mu \to \nu)$ of Eq.~(\ref{eq:mc-rule}) are carefully
determined and given in Fig.~\ref{fig:operation}. (When the chosen
bonds of Fig.~\ref{fig:operation} are located at the ends of the
polymer, the $\eta(\mu\to\nu)$ values will be different from the
values shown in the figure.) With the present MC procedure, we are
able to simulate polymer lengths up to $N=3,000$, which is much
longer than the lengths of several hundreds used in some previous MC
simulations\cite{Wittkop-etal-1995,Dai-etal-2003,Doye-etal-1998}.

\section{Temperature-induced collapse transitions}
\label{sec:temperature}

We begin with the temperature-induced collapse transitions at zero
external force ($f\equiv 0$). We will obtain numerically the phase
diagram of the polymer in the plane spanned by temperature 
$T$ and bending stiffness $\epsilon_{\rm b}$.

\subsection{Flexible polymers ($\epsilon_{\rm b} = 0$)}

First we focus on flexible chains. The equilibrium properties of a
flexible lattice polymer are well-known, therefore this system can
be used to check the validity of our MC simulation method. Let us
denote by $R_N$ the end-to-end distance of the polymer of $N$ bonds.
(In our simulations, we use $R_N$ instead of the gyration radius to
characterize the structural properties of a polymer. $R_N$ behaves
in the same way as the gyration radius \cite{Lifshitz-etal-1978} and
it can be computed with less statistical variance
\cite{Wittkop-etal-1995}). As mentioned in Sec.~\ref{sec:introduction},
there is the following scaling relationship between $R_N$ and the polymer
size $N$:
\begin{equation}
  \label{EQ:power-law}
  \langle R_{N}^{2}\rangle  \sim   N^{2 \nu} \ ,
\end{equation}
where $\langle \cdots \rangle$ means thermal average.
At low temperatures, the polymer is in the compacted globular form,
and the scaling exponent $\nu$ approaches $1/2$ in 2D
\cite{Flory-1967,deGennes-JPL-1975,deGennes-JPL-1978}; at high
temperatures, the configuration of the polymer is reminiscent of a
self-avoiding random walk, and $\nu$ approaches the other limiting
value of $3/4$ [\onlinecite{Nienhuis-1982}]. At the globule-coil phase
transition temperature $T_\theta$, the polymer is in a critical
state, and the scaling exponent $\nu$ takes yet another value
$\nu=\nu_{\rm cr}=4/7$ [\onlinecite{Duplantier-Saleur-1987}]. The scaling
exponent $\nu$ of Eq.~(\ref{EQ:power-law}) in general may have a
weak dependence on the polymer length $N$
\cite{deGennes-1979,Halley-1988,Meirovitch-Lim-1989,Meirovitch-Lim-1990,Torres-etal-1994,Rubio-etal-1995}.
However, at the critical $\theta$ state, $\nu$ will be independent
of $N$. If we plot $\langle R_N^2 \rangle / N^{2 \nu_{\rm cr} }$,
then curves for different lengths $N$ should intersect at the
$\theta$ temperature.

\begin{figure}[b]
    \includegraphics[angle=0, width=0.6\linewidth]{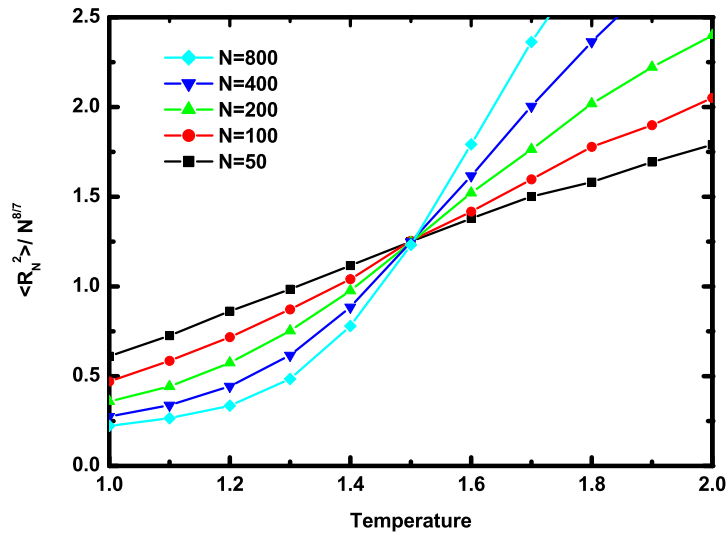}
  \caption{\label{fig:scale_S00_F00_T} 
   (Color Online) The relationship between $\langle R_N^2 \rangle / N^{8/7}$ and temperature
    $T$ for a flexible
    lattice polymer of different lengths $N$.
    Different curves intersect at the $\theta$ temperature
    of $T_\theta\approx 1.50$.
  }
\end{figure}

Figure \ref{fig:scale_S00_F00_T} shows the relationship between
$\langle R_N^2 \rangle/ N^{8/7}$ and temperature for chains of
different lengths $N$. Different curves indeed intersect at the same
point when $T= T_\theta \approx 1.50$. 
Therefore, the present MC method is able to correctly predict 
the critical exponent of $\nu_{\rm cr}=4/7$ for a
2D flexible polymer. The predicted $T_\theta$ value is also in agreement with
the value of $T_\theta = 1.54 \pm 0.05$ as given by Ref.~[\onlinecite{Seno-Stella-1988}].

\begin{figure}
  \includegraphics[angle=0, width=0.6\linewidth]{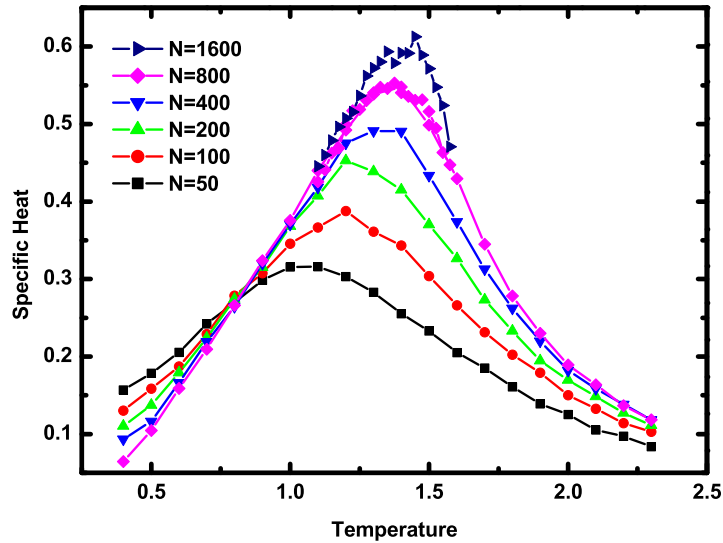}
  \caption{\label{fig:SpecificHeat_S00_F00_T} 
    (Color Online) The specific  heat $c(T)$ for a 2D flexible lattice polymer of
    different lengths $N$.
  }
\end{figure}

\begin{figure}
  \includegraphics[width=0.6\textwidth]{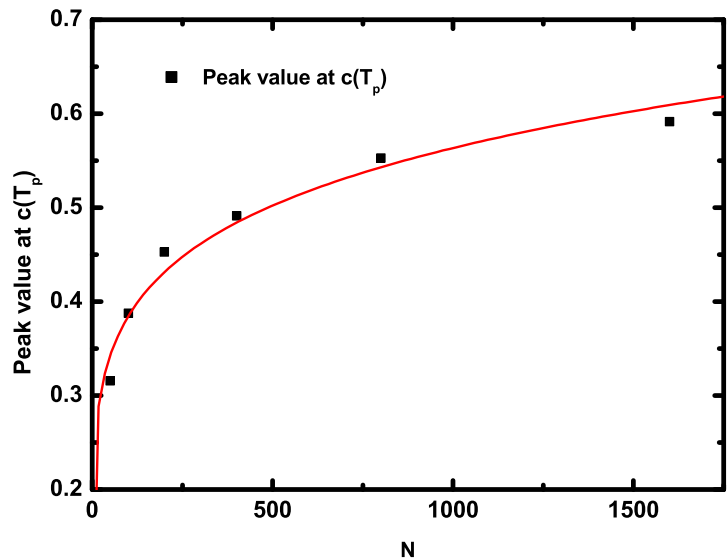}
  \caption{\label{fig:FaiS0F0} 
    The relation between the peak values $c(T_p)$ of the
    specific heat and chain length for a flexible polymer
    ($\epsilon_{\rm b}=0$). The solid line is the fitting curve
    $c(T_p)= b N^\alpha$ with $b=0.179\pm0.017$ and $\alpha=0.166\pm
    0.015$.
  }
\end{figure}

When there is no external force, the specific heat $c(T)$ of the system as a function of $T$ can be
obtained through $c(T)= \beta^{2} \bigl(\langle E^{2}\rangle-\langle E\rangle ^{2} \bigr)/ N $ and is shown in
Fig.~\ref{fig:SpecificHeat_S00_F00_T}. As the chain length $N$ increases,
surface effect (which reduces the total number of contacts) of the globular
phase becomes less important, therefore the temperature $T_p$ at which  $c(T)$ 
reaches maximum increases with $N$. The MC simulation data suggest
that $T_p$  approaches the critical temperature $T_\theta$ according to the
following formula $T_p = T_\theta [1-(c_0 / N)^s ] $ with $c_0 = 3.0 \pm 1.0 $
and $s = 0.44 \pm 0.04$. The peak value $c(T_p)$ of the specific heat 
diverges with chain length $N$ according to $c(T_p) \propto N^\alpha$ with 
$\alpha = 0.166 \pm 0.015$ (see Fig.~\ref{fig:FaiS0F0}). From this finite-size scaling
behavior we estimate the
crossover exponent \cite{Brak-etal-1993} $\phi \equiv 1/(2-\alpha)$  to be
$\phi \approx 0.545$.  This value is in agreement with the
value of $\phi = 0.52 \pm  0.07$ given by Ref.~[\onlinecite{Seno-Stella-1988}].

Taking together, the simulation results of this subsection, in confirmation to various earlier efforts, 
predict a continuous globule--coil phase-transition for a self-attracting flexible 2D polymer.

\subsection{Semiflexible polymers}
\label{subsection:semiflexible-T}

\begin{figure}[t]
  \includegraphics[width=0.3\textwidth]{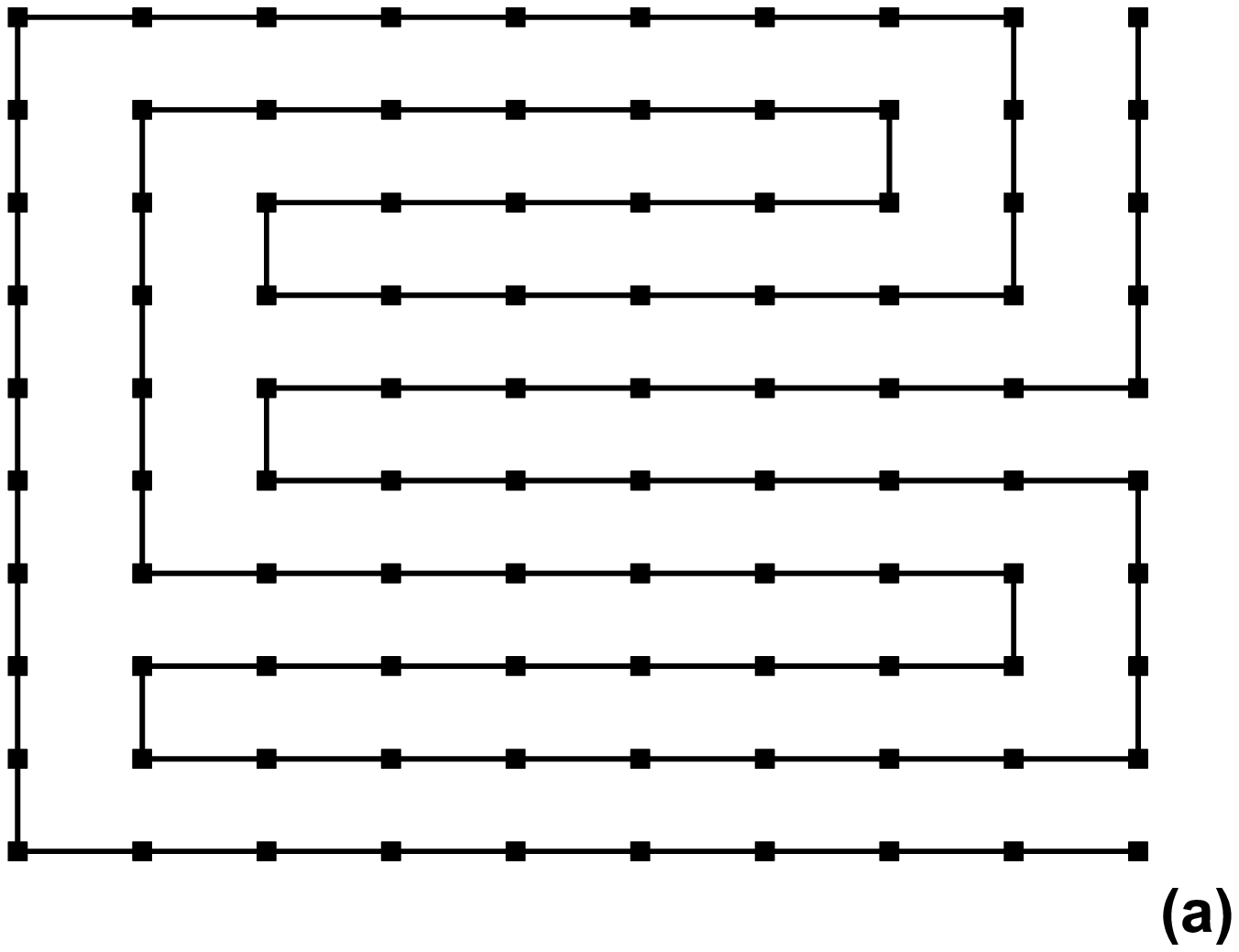}
  \includegraphics[width=0.3\textwidth]{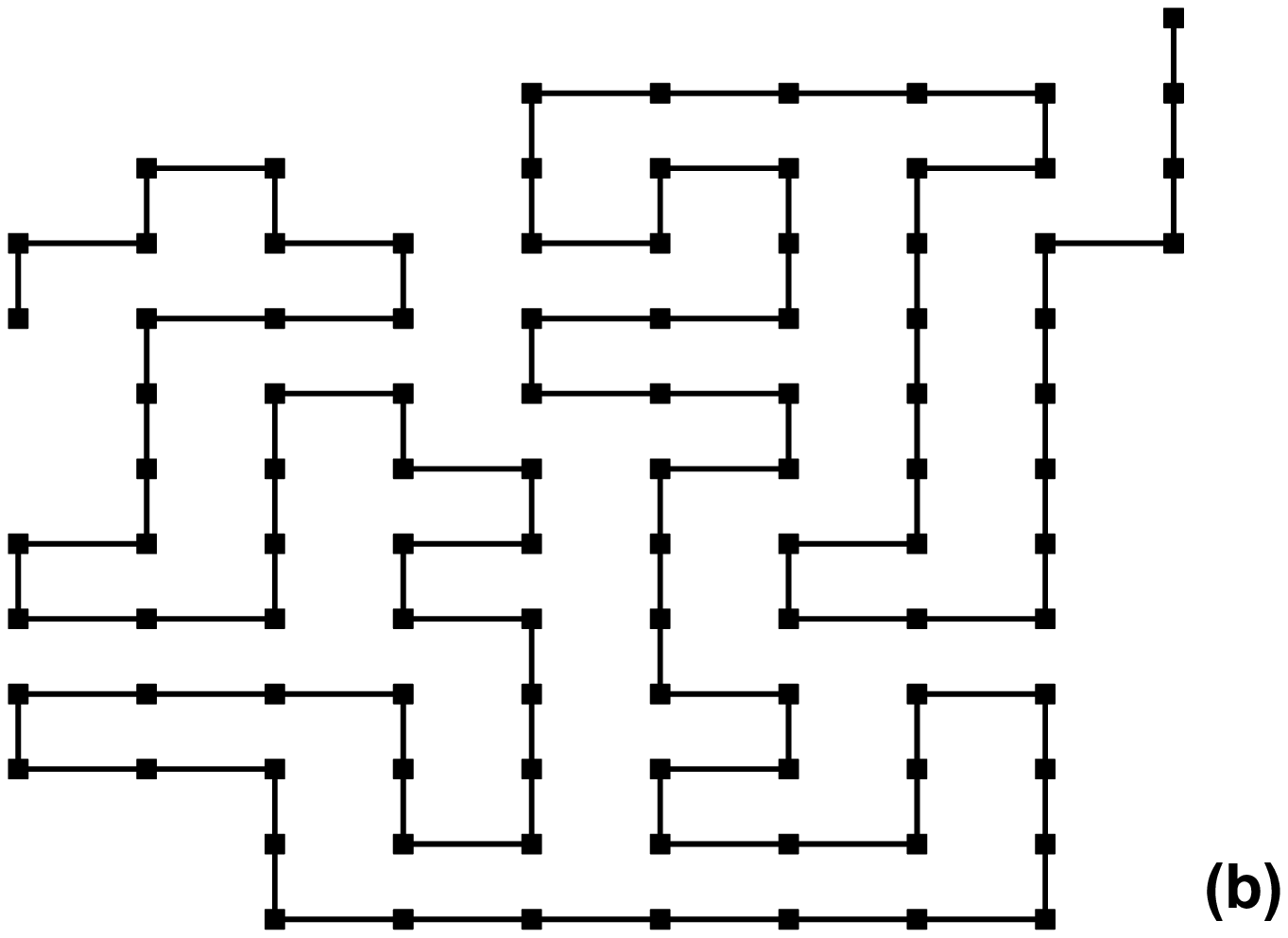}
  \includegraphics[width=0.3\textwidth]{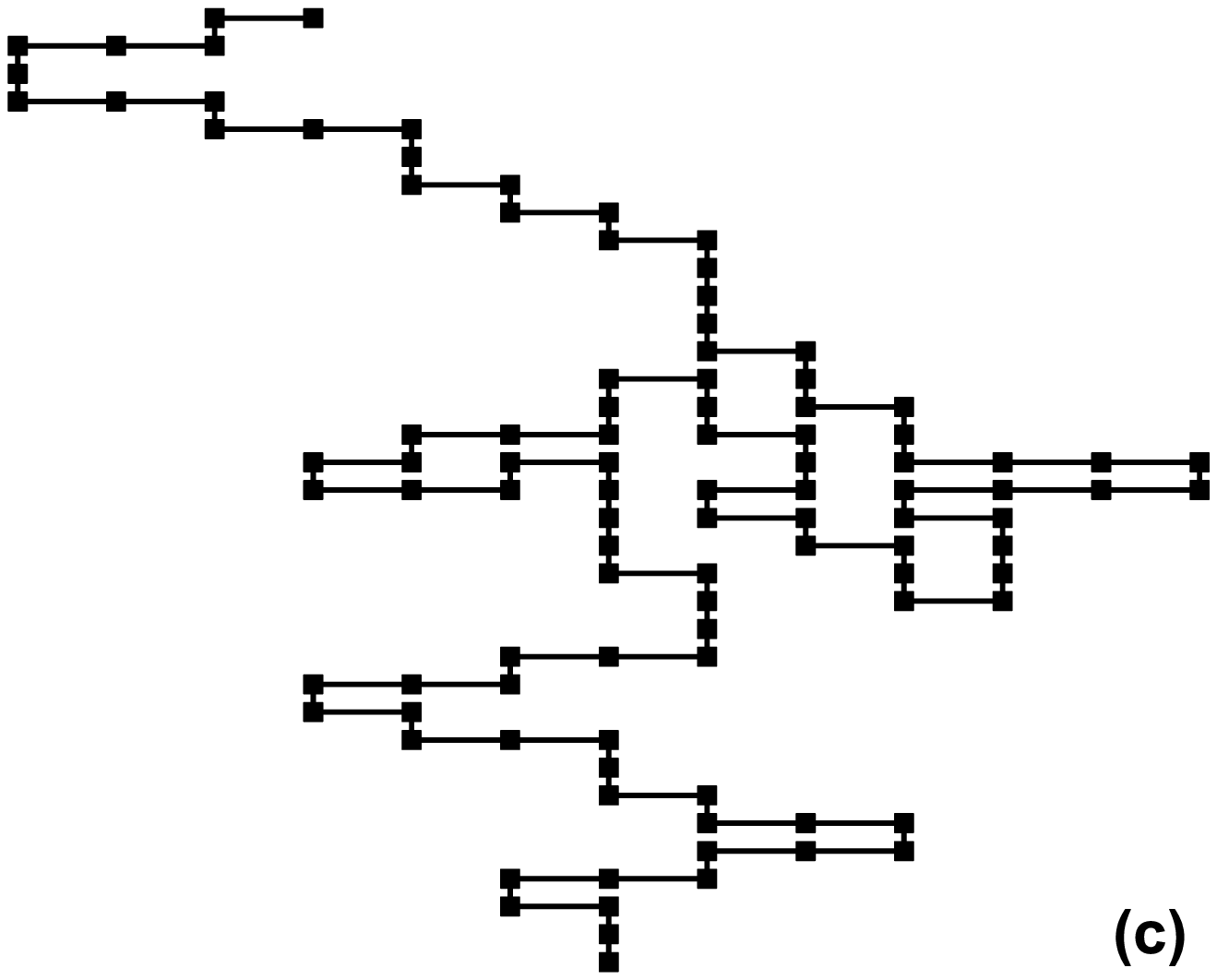}
  \caption{\label{fig:solid-1H} 
    Typical configurations of a semiflexible polymer
    in the crystal (a), disordered globule (b), 
    and extended random coil (c) phase.
  }
\end{figure}

\begin{figure}[b]
  \includegraphics[angle=0,width=0.6\linewidth]{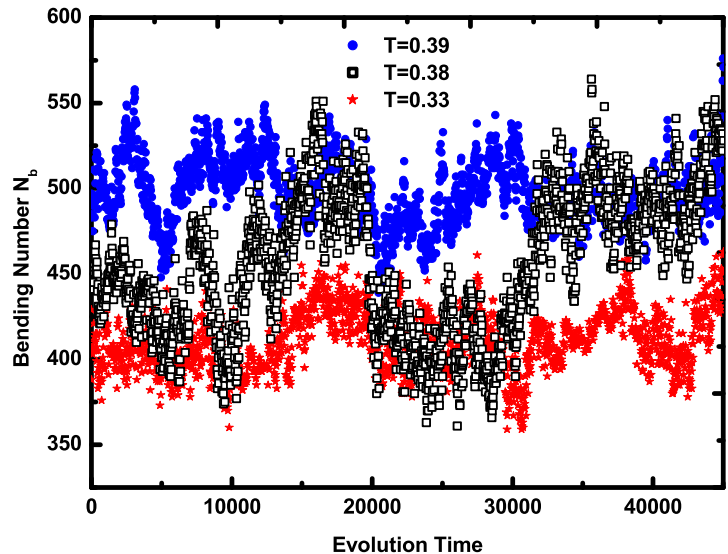}
  \caption{\label{fig:s-g} 
    (Color Online) Evolution of the configuration of a semiflexible polymer with 
    bending stiffness
    $\epsilon_{\rm b}=0.3$ and length $N=2,000$.
    The vertical axis is the number of bends $N_{\rm b}$
    in the configuration, and the horizontal axis is the evolution
    time in terms
    of MC steps. (At time zero of the figure the system has already 
    reached
    equilibrium.)
    The temperature is $T=0.33$ (red stars),
    $0.38$ (black open squares), and $0.39$ (green filled squares).
  }
\end{figure}

When bending stiffness $\epsilon_{\rm b}$ of the polymer is
positive, the polymer's configuration is also influenced by bending
energies. At very low temperatures entropic effect is not important,
then the polymer will favor those compact configurations which have
the maximal number of monomer-monomer contacts and the minimal
number of bends. These are highly ordered crystal configurations as
illustrated in Fig.~\ref{fig:solid-1H}(a).  In such a crystal
configuration
\cite{Zwanzig-Lauritzen-1968,Zwanzig-Lauritzen-1970,Kuznetsov-etal-1996}
long segments of the polymer stack onto each other, and the total
number of bends $N_{\rm b}$ scales only sub-linearly with chain
length $N$.
 Let us define two order parameters $n_{\rm b}$ (the density of bends) and
$n_{\rm c}$ (the density of contacts) as
\begin{eqnarray}
n_{\rm b} & = & \frac{N_{\rm b}}{N} \ ,  \label{eq:n_b} \\
n_{\rm c} & = & \frac{N_{\rm c}}{N}\ .   \label{eq:n_c}
\end{eqnarray}
For a polymer in a crystal conformation, we have $\lim_{N\to \infty}
n_{\rm b} = 0$ and $\lim_{N \to \infty} n_{\rm c}$ of order unity.
The overall shape of a polymer crystal may  be a
rectangle rather than a square, the aspect ratio of which depending
on $\epsilon_{\rm b}$. When temperature is elevated, the entropy
effect becomes more and more important. At certain point the crystal
order of the polymer will be destroyed. For $\epsilon_{\rm b}$ being
small, when crystal order disappears the monomer-monomer contacting
interactions may still be very significant. Then the polymer will
take disordered globular shapes (Fig.~\ref{fig:solid-1H}(b)). In  a
disordered globular configuration, both the order parameters $n_{\rm b}$
and $n_{\rm c}$ are of order unity. When temperature is further increased, the
globular configuration will  again become unstable, and the polymer
will transit to the phase of swollen random coil
(Fig.~\ref{fig:solid-1H}(c)).

The crystal-globule structural transition is a first-order
phase-transition in the thermodynamic limit of $N\to \infty$,
similar to the solid-liquid transition of water at $0$$^\circ$C.
This is demonstrated in Fig.~\ref{fig:s-g} for the evolution of the
total bending number $N_{\rm b}$ of a polymer with length $N=2,000$
and bending stiffness $\epsilon_{\rm b}=0.3$. At temperature
$T=0.33$, $N_{\rm b}$ fluctuates around $N_{\rm b}^{(1)} \approx
410$. At $T=0.39$, $N_{\rm b}$ fluctuates around the value of
$N_{\rm b}^{(2)}\approx 510$. While at the intermediate temperature
$T=0.38$, $N_{\rm b}$ jumps between these two values $N_{\rm
b}^{(1)}$ and $N_{\rm b}^{(2)}$, indicating that the system is
bistable. On the other hand, at each of these temperatures, the
histogram of the number of contacts $N_{\rm c}$  has only one peak.
The crystal-globule transition temperature increases with the
bending stiffness $\epsilon_{\rm b}$ of the polymer.

\begin{figure}
  \includegraphics[angle=0,width=0.6\linewidth]{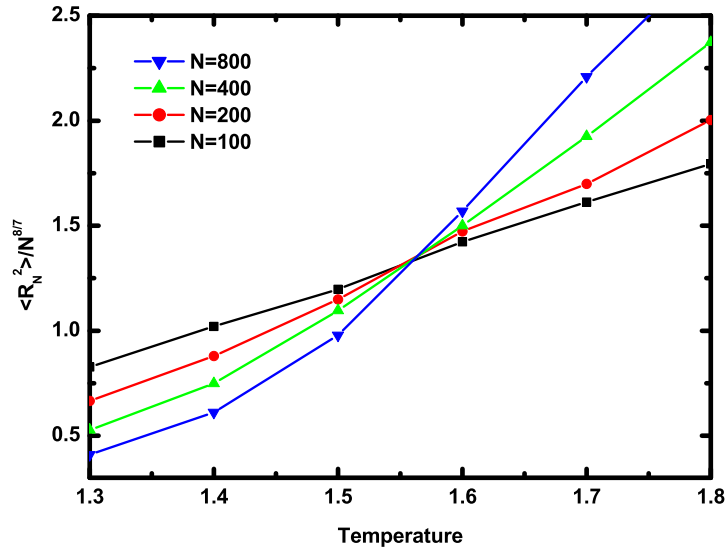}
  \caption{\label{fig:scale_S06_F00_T} 
    (Color Online) The relationship between $\langle R_N^2 \rangle / N^{8/7}$ and
    temperature
    $T$ for a semiflexible
    polymer of different lengths $N$. The bending stiffness is
    $\epsilon_{\rm b}=0.6$.
    Different curves intersect at the $\theta$ temperature
    of $T_\theta\approx 1.55$.
  }
\end{figure}

\begin{figure}
  \includegraphics[width=0.6\linewidth]{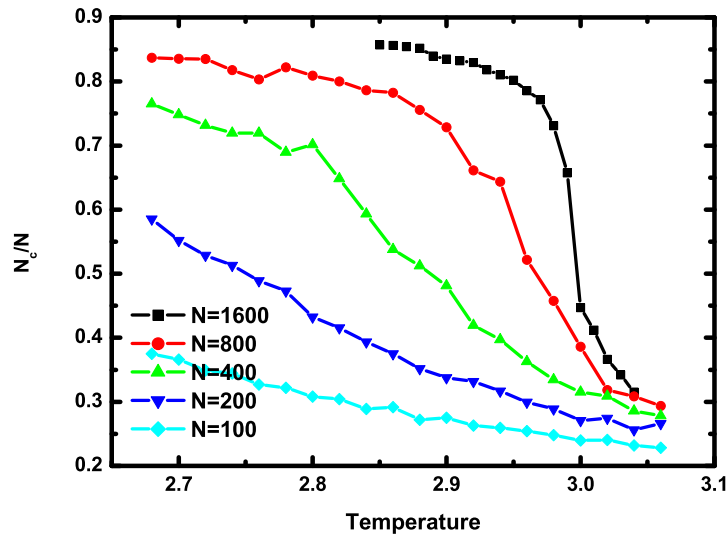}
  \caption{\label{fig:Nc_S5F0}  
    (Color Online) The order parameter $n_{\rm c}$ as a function of temperature $T$
    for a semiflexible chain with $\epsilon_{\rm b}=5.0$ and different lengths $N$.
  }
\end{figure}

When the temperature is further increased to $T\approx 1.5$, a
second structural transition takes place. This globule-coil
transition is second-order, as can be inferred from
Fig.~\ref{fig:scale_S06_F00_T}. This figure shows the value of
$\langle R_N^2 \rangle/ N^{8/7}$ as a function of temperature for a
semiflexible polymer of bending stiffness $\epsilon_{\rm b}=0.6$ and
different lengths $N$ (similar results were obtained for
$\epsilon_{\rm b}=0.3$ and $\epsilon_{\rm b}=1.0$).
Figure~\ref{fig:scale_S06_F00_T} is qualitatively the same as
Fig.~\ref{fig:scale_S00_F00_T} of a flexible polymer. The only
quantitative difference is that the transition $\theta$ temperature
is slightly increased. At the globule-coil transition point, the
scaling exponent $\nu_{\rm cr}$ of the semiflexible polymer is the
same as that of a flexible polymer. Therefore, bending stiffness
does not influence the scaling behavior of a polymer in its critical
$\theta$ state. The globule-coil transition temperature is also
not very sensitive to the bending stiffness.

In the crystal or the disordered globular phase, the size of the
lattice polymer scales with chain length in a square-root law,
i.e., $R_N^2 = c_1 N$ with $c_1 \approx 0.6$. The value of the prefactor $c_1$
has no obvious change
at the crystal-globule transition temperature; it is also 
insensitive to the bending stiffness $\epsilon_{\rm b}$. When the temperature 
approaches or goes beyond the globule-coil transition temperature $T_\theta$,
however, $c_1$ increases quickly with $T$ (because the
square-root scaling law no-longer holds).

\begin{figure}
  \includegraphics[width=0.45\linewidth]{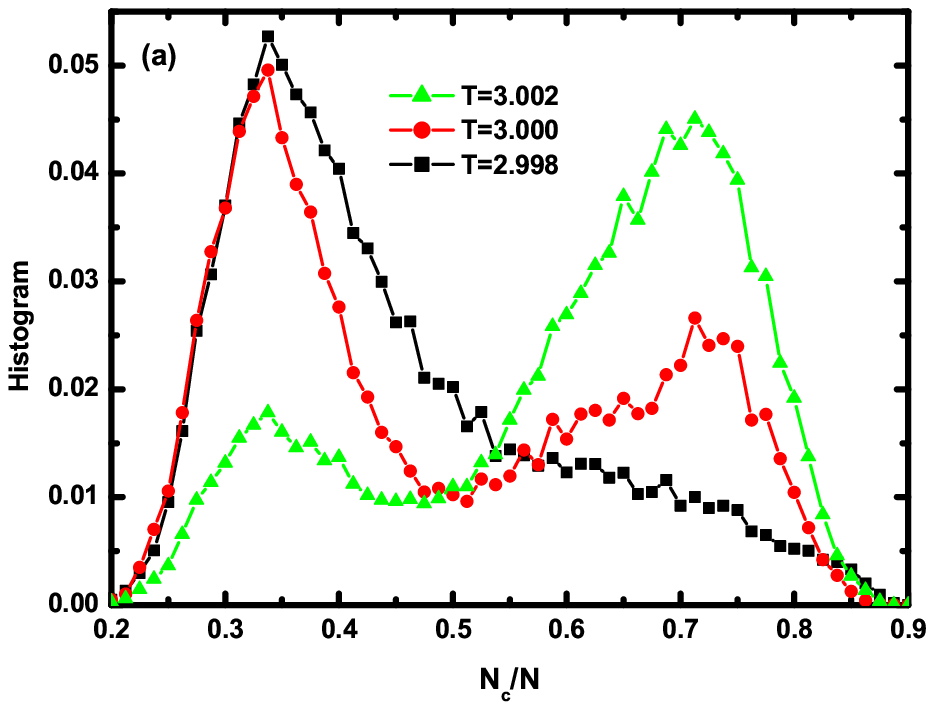}
  \includegraphics[width=0.45\linewidth]{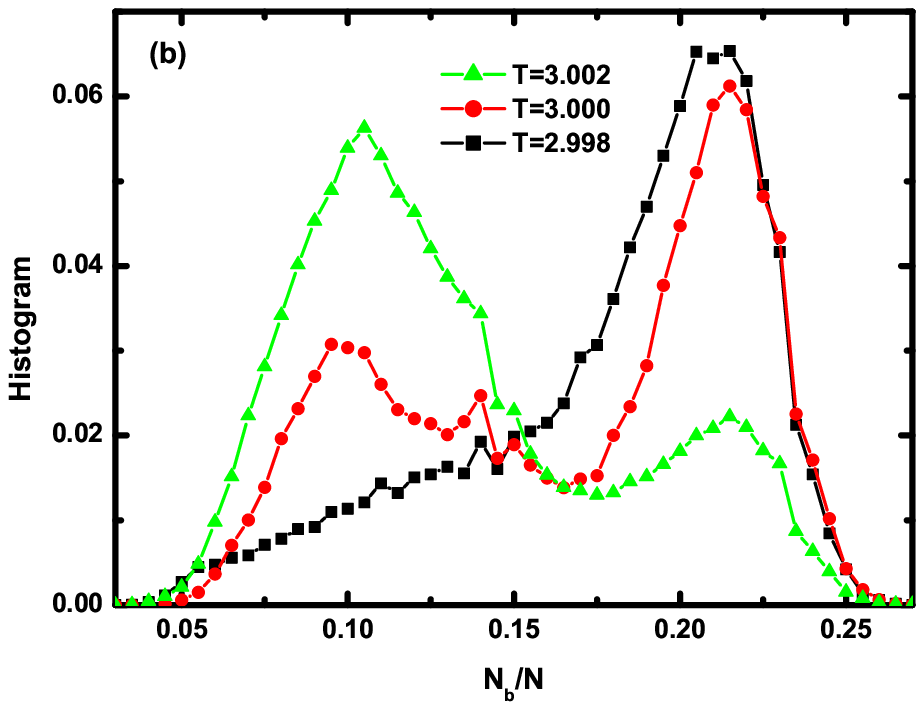}
  \caption{\label{fig:Contact-Distribution-H08-S50-F00-T} 
    (Color Online) Distribution of contact number $N_{\rm c}$ (a) and bending number
    $N_{\rm b}$ (b) for a
    semiflexible polymer with bending stiffness $\epsilon_{\rm b}=5.0$ and length $N=1,600$.
    Different curves correspond to slightly different temperatures.
  }
\end{figure}

\begin{figure}
  \includegraphics[angle=0,width=0.6\linewidth]{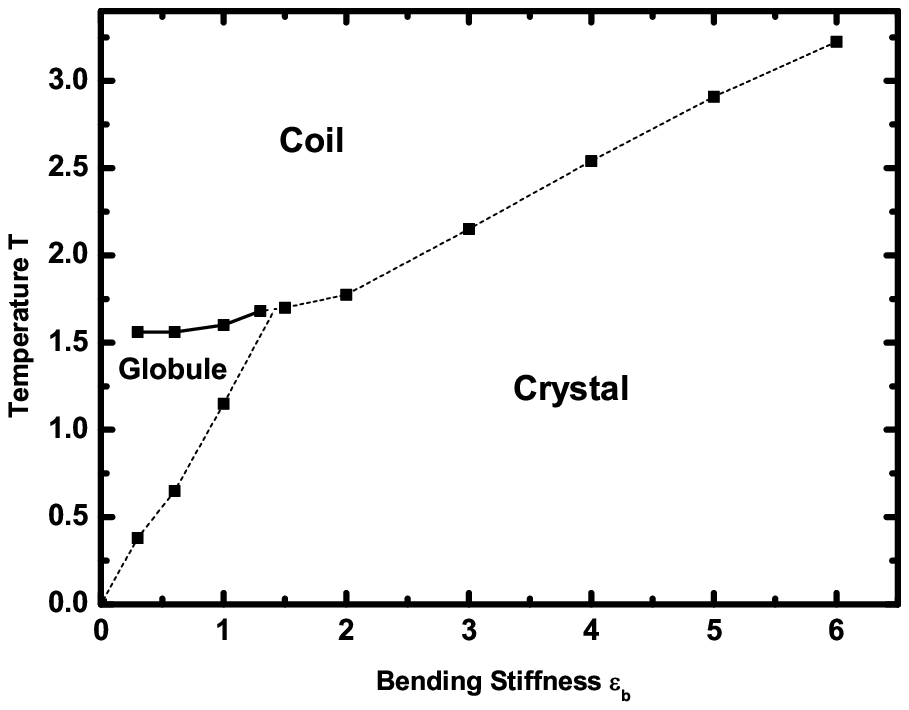}
  \caption{\label{fig:temperature-delta-phase-diagram} 
    The zero-force phase-diagram of the 2D lattice polymer. Square symbols are simulation results. 
    The crystal-globule transition and the crystal-coil
    transition is first-order (as indicated by dashed boundary lines) 
    in the thermodynamic limit, while the globule-coil transition is
    continuous (solid boundary line).
  }
\end{figure}

When the bending stiffness of the semiflexible polymer
exceeds a threshold value of $\epsilon_{\rm b} \approx 1.5$, the
intermediate disordered globule phase disappears. The polymer will
exists only in two phases, either the low-temperature crystal phase
or the high-temperature swollen coil phase, and the structural
transition between these two phases is first-order. For a
semiflexible polymer with $\epsilon_{\rm b}=5$ and length $N$, Fig.~\ref{fig:Nc_S5F0}
shows the relationship between the density of contacts $n_{\rm c}$
and temperature $T$. As $N$ becomes sufficiently
large (e.g., $N=1,600$) we notice a
dramatic drop of $n_{\rm c}$ at $T \approx 3.0$, indicating the
existence of a globule-coil transition.  For $T$ close to this
transition temperature, 
Fig.~\ref{fig:Contact-Distribution-H08-S50-F00-T}(a) and
Fig.~\ref{fig:Contact-Distribution-H08-S50-F00-T}(b)
report the distributions of the polymer's contacting 
number $N_{\rm c}$ and bending number $N_{\rm b}$.
When the temperature $T$ is close to
$T=3.0$, there are two peaks in both probability distributions,
indicating the system is bistable (both in terms of
$N_{\rm c}$ and $N_{\rm b}$) in this temperature range. The
positions of these two peaks in the distribution of $N_{\rm c}$ and
that of $N_{\rm b}$ remain almost fixed when the temperature
changes, but their weights changes. As temperature increases, the
system has higher probability to be in the coil phase. 
We can define the crystal-coil transition temperature 
as the temperature at which
the system has equal probability to be in the crystal and the coil
phase. This transition is between the crystal phase and the coil
phase, because both the order parameters $n_{\rm c}$ and $n_{\rm b}$ will
jump in the thermodynamic limit of $N\to \infty$ rather than
just one of them.

We have performed the same analysis for other values of the bending
stiffness $\epsilon_{\rm b}$.
At zero external force, the complete phase-diagram of a 2D lattice
polymer is shown in
Fig.~\ref{fig:temperature-delta-phase-diagram}.
When $\epsilon_{\rm b} < \epsilon_{\rm b}^* \approx 1.5$, the
polymer has two structural phase-transitions as temperature changes,
a first-order crystal-globule transition followed by a
higher-temperature continuous globule-coil transition. When
$\epsilon > \epsilon_{\rm b}^*$, the polymer will transit directly
from the crystal phase to the coil phase. In
Ref.~[\onlinecite{Zhou-etal-2006}], it was predicted based on a partially
directed 2D lattice model that, the collapse transition is
first-order as long as the bending stiffness is positive. This
apparent contradiction between Ref.~[\onlinecite{Zhou-etal-2006}] and the
present work is easy to understand. In the partially directed
polymer model, the compact phase, since it is highly ordered,  is
not the disordered globule phase but rather the crystal phase.
Because of the additional constraint of partial directness, the
entropy of the compact phase is suppressed compared to that of the
disordered globule phase. The simulation results of
Fig.~\ref{fig:temperature-delta-phase-diagram} also suggests that
the crystal-coil transition temperature increases almost linearly
with the bending stiffness $\epsilon_{\rm b}$, in consistence with
the earlier work of Doniach and co-workers on
3D lattice polymers \cite{Doniach-etal-1996}.

\section{Force-induced phase-transitions}
\label{sec:force}

With the zero-force phase-diagram
Fig.~\ref{fig:temperature-delta-phase-diagram} being obtained for a
2D lattice polymer, we now proceed to study its force-induced
structural transitions. Initially the polymer is in a
low temperature crystal or disordered globule phase.  An external
force along the $x$-direction is applied to the free end of the
polymer, and the $x$-component of the polymer's extension, $X_N$,
the contacting number $N_{\rm c}$, the bending number $N_{\rm b}$
are recorded during the stretching.

\subsection{The globule-coil transition}

For a 2D lattice polymer with bending stiffness $\epsilon_{\rm b} <
1.5$, we can choose an appropriate temperature value (say $T=1.2$)
such that, at zero external force, the polymer is in the disordered
globular phase. At this temperature, when the external force $f$
becomes sufficiently large, the globular shape of the polymer will
be destroyed. The polymer will be in a highly extended coil
configuration. As has been predicted by several earlier theoretical
work \cite{Grassberger-Hsu-2002,Zhou-etal-2006}, in the
thermodynamic limit of $N \to \infty$, a globule-coil transition
will take place at certain critical force $f_{\rm cr}(T)$. In
agreement with Ref.~[\onlinecite{Grassberger-Hsu-2002}], we find
this 2D
force-induced globule-coil transition is a continuous transition.

\begin{figure}
  \includegraphics[width=0.6\linewidth]{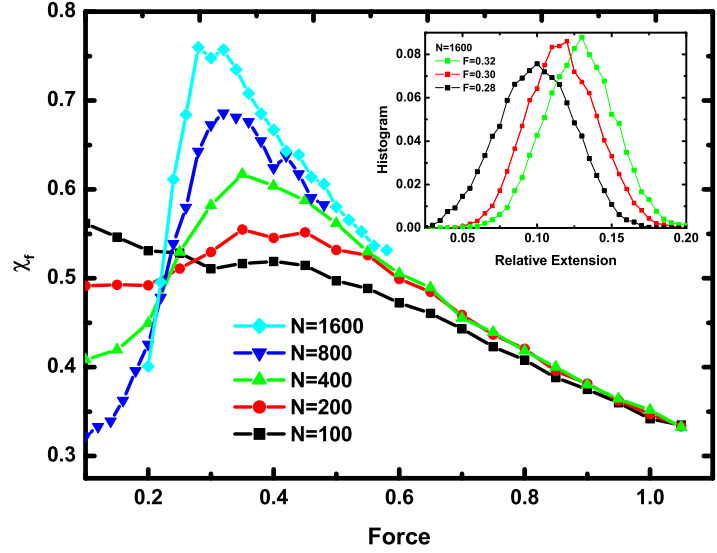}
  \caption{\label{fig:chi-F} 
    (Color Online) The extension susceptibility $\chi_{f}$ as a function of external force
    $f$ at $T=1.2$ 
    for a polymer with $\epsilon_{\rm b}=0$ and
    chain length $N$. Insert shows the distributions of normalized mean extension
    $\langle X_N \rangle / N$ for the polymer system with $N=1,600$ near the globule-coil
    transition.
  }
\end{figure}

The continuous nature  of the force-induced globule-coil transition
is checked by investigating the fluctuation of the
elongation $X_N$. Figure~\ref{fig:chi-F} shows how the extension
susceptibility $\chi_f$ of a flexible polymer system ($\epsilon_{\rm b}=0$) at
$T=1.2$ 
changes with force $f$. The
extension susceptibility is defined by
\begin{equation}
  \chi_f \equiv  \frac{1}{N}
  \frac{ \partial \langle X_N \rangle }{\partial f} = \beta \bigl[ \langle X_N^2 \rangle - \langle X_N \rangle^2 \bigr] / N \ .
  \label{eq:chi-F}
\end{equation}
Figure~\ref{fig:chi-F} demonstrated that, as the chain length $N$
increases, the peak of $\chi_f$ becomes more pronounced (eventually
it will approach infinity). This behavior is very similar to
the divergence of the specific heat $c(T)$ at the temperature-induced
globule-coil transition (see Sec.~\ref{sec:temperature}).
By finite-size scaling and extrapolating to the case of $N =
\infty$ we obtain a transition force $f_{\rm cr} \approx  0.3$ for the system
(an more precise estimate of the transition is reported below).
The inset of  Fig.~\ref{fig:chi-F} 
shows the distribution of the extension of the system at different
external forces. When the external force is close to the
transition force $f_{\rm cr}$, the extension distribution has 
only one peak which moves with
force. This behavior further confirms the continuous nature
of the structural transition.

\begin{figure}
  \includegraphics[width=0.6\textwidth]{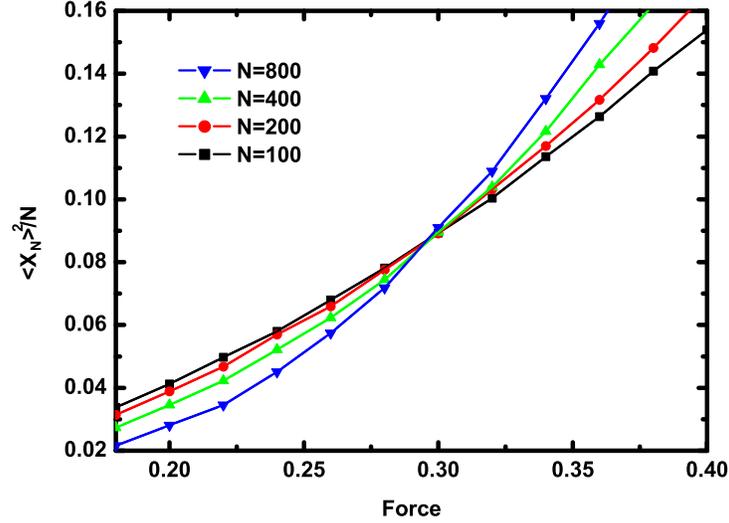}
  \caption{\label{fig:Interscetion2_S00_T12_F} 
    (Color Online)
    The value of $\langle X_N \rangle^2 / N^{1.74}$ as a function of external force $f$
    for a flexible polymer ($\epsilon_{\rm b}=0$) of different lengths $N$.
    Different curves intersect at $f\approx 0.295$.
  }
\end{figure}

Under an external force, we can write down the following effective scaling
relation between the mean extension of the polymer and the chain length
$N$:
\begin{equation}
  \langle X_N \rangle^2  \sim N^{\mu} 
  \label{eq:X-N-scaling}
\end{equation}
with effective exponent $\mu$. 
In the globule phase, the polymer is in an elliptic shape and $\mu$ 
in Eq.~(\ref{eq:X-N-scaling}) is
less than $2$ but larger than $1$. $\mu$ may
increase with force $f$ and reach a limiting value $\mu_{\rm cr}$ at the
globule-coil transition point. When $f$ further increases and the polymer
is in the coil phase, the extension of the polymer scales linearly with 
chain length $N$, i.e., $\mu=2$. As in Sec.~\ref{sec:temperature}, we
expect that, at the globule-coil transition point, the critical exponent
$\mu_{\rm cr}$ will be independent of chain length $N$. Then if
$\langle X_N \rangle^2 / N^{\mu_{\rm cr}}$ is
plotted as a function of $f$, curves corresponding to different chain length
$N$ will intersect at the same point $f_{\rm cr}$. 
Figure \ref{fig:Interscetion2_S00_T12_F} demonstrates that this
is indeed the case, and from this plot we estimate that $f_{\rm cr}=0.295$ and
$\mu_{\rm cr}=1.74$ for a flexible polymer at $T=1.2$. This value of $f_{\rm cr}$
is in agreement with Ref.~[\onlinecite{Grassberger-Hsu-2002}], which 
predicted $f_{\rm cr} \approx 0.30$. To further validate this
method of determining the critical force. we have used it
to determine the critical force for a flexible 2D partially directed
polymer at $T=0.59$ as studied in Ref.~[\onlinecite{Zhou-etal-2006}]. We 
obtained that the critical force $f_{\rm cr}=0.45$ (very close to the
exactly known value $0.46$) and $\mu_{\rm cr}=1.2$.

\begin{figure}
  \includegraphics[width=0.6\textwidth]{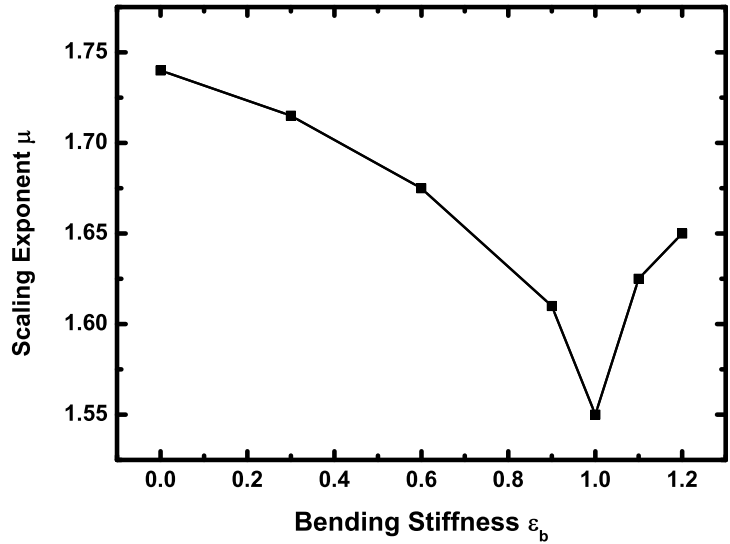}
  \caption{\label{fig:Scaling_Stiff_Force} 
    The critical scaling exponent $\mu_{\rm cr}$ at the globule-coil transition
    of Eq.~(\ref{eq:X-N-scaling}) as a
    function of the bending stiffness $\epsilon_{b}$ at temperature $T=1.2$.
  }
\end{figure}

\begin{figure}
  \includegraphics[width=0.6\linewidth]{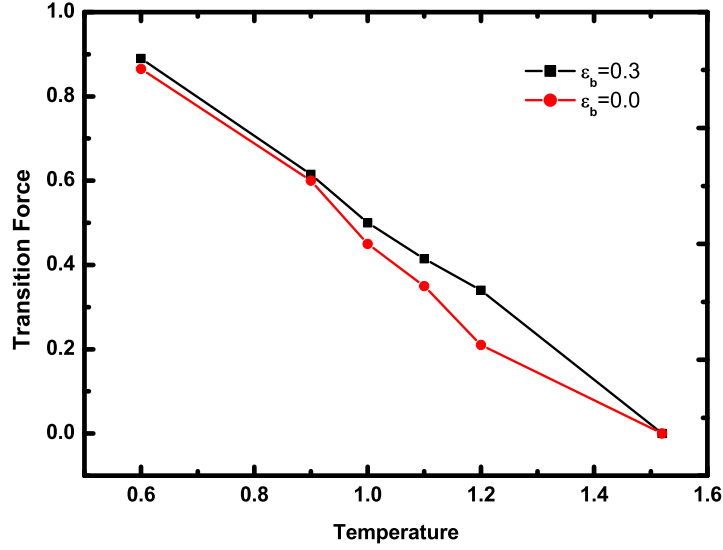}
  \caption{\label{fig:Fc_Tp} 
    The globule-coil transition force $f_{\rm cr}(T)$ as a function of temperature $T$.
  }
\end{figure}

For $T=1.2$ we have determined the critical exponent
$\mu_{\rm cr}$  for 2D lattice polymers with
different bending stiffness $\epsilon_{\rm b}$. The critical
exponent $\mu_{\rm cr}$ is not a constant. It drops from the value of $\mu_{\rm cr} =
1.74$ for $\epsilon_{\rm b}=0$ to $\mu_{\rm cr} = 1.55$ for $\epsilon_{\rm   b}=1.0$ 
(see Fig.~\ref{fig:Scaling_Stiff_Force}). Intuitively this
observation is natural to understand. At the force-induced
globule-coil phase-transition point, the polymer takes the shape of
an elongated ellipse \cite{Grassberger-Hsu-2002}. The bending
stiffness $\epsilon_{\rm b}$ affects the shape of this ellipse and
therefore affects the critical exponent $\mu_{\rm cr}$. On the other hand, to
derive an explicit expression for the relationship between $\mu$ and
$\epsilon_{\rm b}$ is a hard task. The critical exponent $\mu_{\rm cr}$ may also depend
on temperature.

The globule-coil transition force $f_{\rm cr}(T)$ decreases with temperature almost
linearly and vanishes at the $\theta$-temperature $T_\theta$, i.e., 
$f_{\rm cr}(T) \sim (T_\theta - T)$ for $T < T_\theta$ (see Fig.~\ref{fig:Fc_Tp}).
This linear relationship holds both for flexible and semiflexible
chains.

\subsection{The crystal-coil transition}

\begin{figure}
  \includegraphics[width=0.45\textwidth]{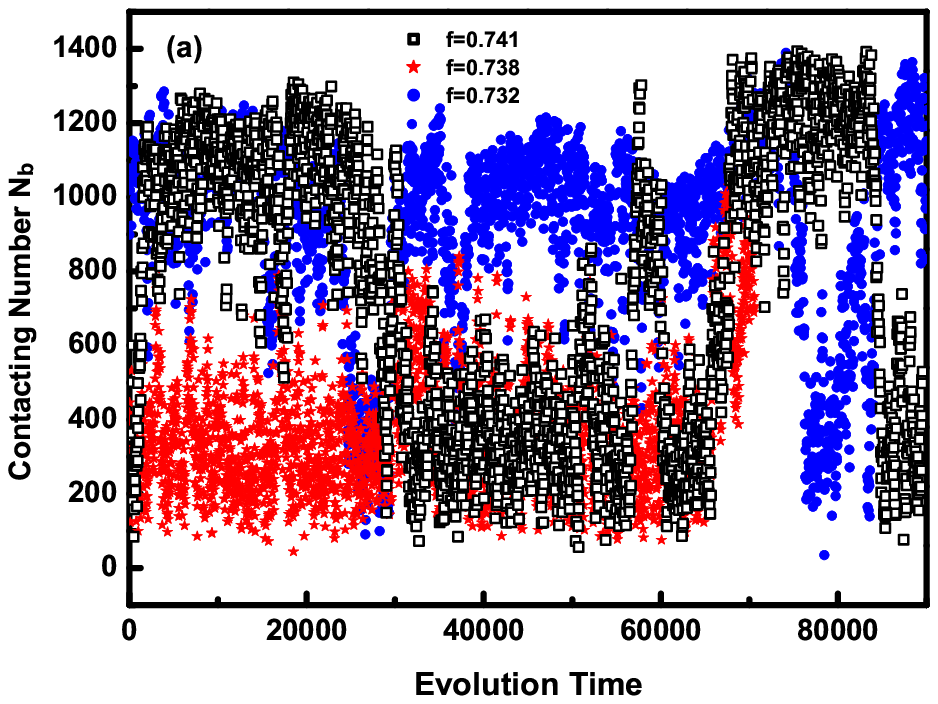}
  \includegraphics[width=0.45\linewidth]{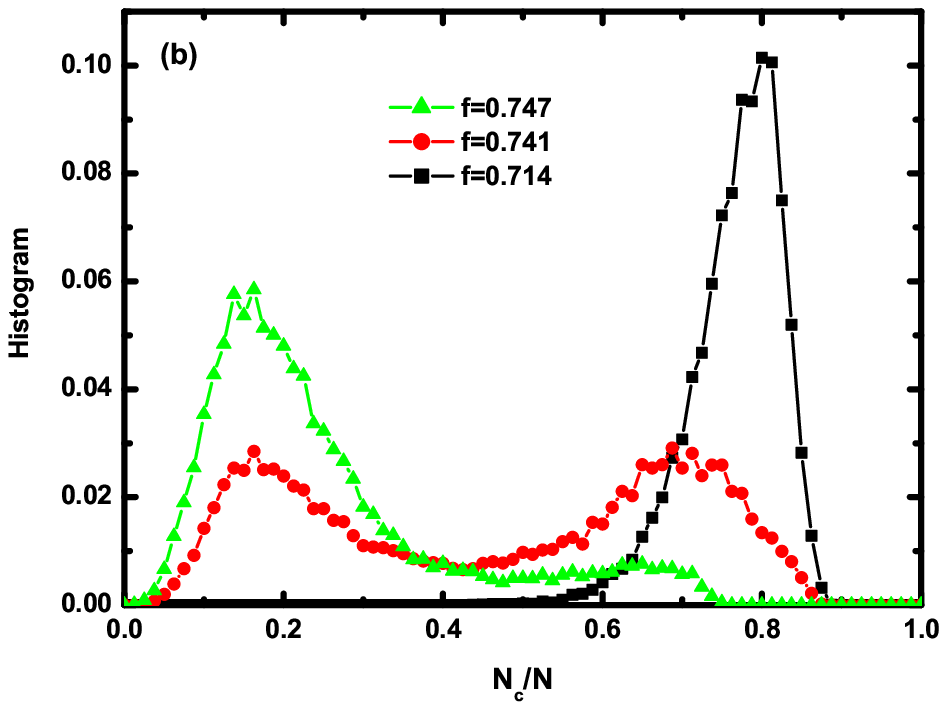}
  \caption{ \label{fig:Contact_Distribution_H16_S50_T25} 
    (Color Online) The time evolution (a) and probability histogram (b) of the
    contacting number $N_{\rm c}$ of a  polymer with $N=1,600$ and $\epsilon_{\rm b}=5.0$.
    Different data sets correspond to different external force $f$, while
    the temperature is fixed at $T=2.5$.
  }
\end{figure}

At low temperature and zero force, a considerably stiff semiflexible
polymer is in the crystal phase. We now investigate the
force-induced crystal-coil transition by MC simulation. Since at
$f=0$, the temperature-induced crystal-coil transition is already a
first-order transition, we expect the force-induced transition
should also be first-order. This is confirmed by MC simulation
results. Figure~\ref{fig:Contact_Distribution_H16_S50_T25}(a) shows
the evolution of the total contacting number $N_{\rm c}$ of a
polymer of length $N=1,600$ and bending stiffness $\epsilon_{\rm
b}=5.0$ at $T=2.5$. At the three different force values, the
contacting number jumps between the coiled shapes of $N_{\rm c}
\approx 300$  and the crystal shapes of $N_{\rm c}\approx 1,000$.
The histogram of the contacting number as shown in
Fig~\ref{fig:Contact_Distribution_H16_S50_T25}(b) has two
well-defined peaks for force $f \approx 0.74$. When $f$ deviates
remarkably from this value, however, one of the peaks in the
contacting number histogram disappears, indicating the system is
either in the crystal phase or in the coil phase.

\begin{figure}
  \includegraphics[width=0.6\textwidth]{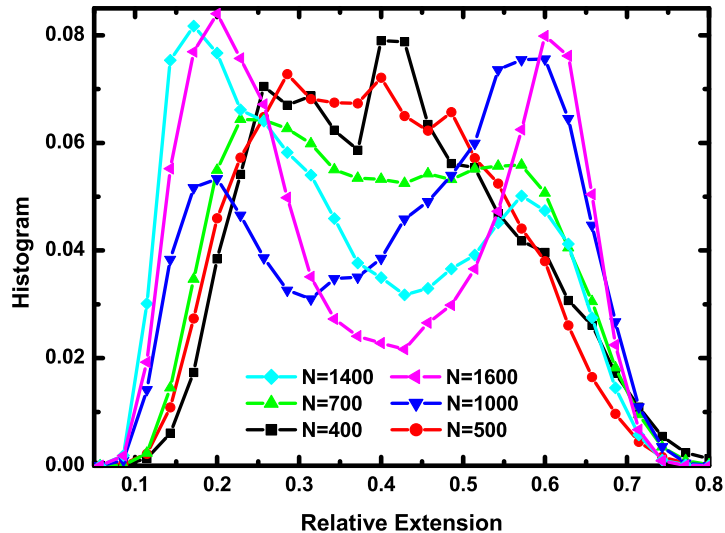}
  \caption{ \label{fig:fcs_S50T25} 
    (Color Online) Distribution of normalized mean extension $\langle X_N \rangle / N$
    for a polymer with
    $\epsilon_{\rm b}=5.0$ at $T=2.5$. The external force is set to the value at which the
    susceptibility $\chi_f$ reaches maximum, with $f=0.68$, $0.67$, $0.70$, $0.72$, $0.73$
    and $0.74$ for $N=400$, $500$, $700$, $1,000$, $1,400$ and $1,600$ respectively.
  }
\end{figure}

Although the crystal-coil transition is a first-order phase-transition,
if the chain length $N$ is too small, the jump in the order parameter
$n_{\rm c}$ or $n_{\rm b}$ can not be observed. This is because  the
correlation length at the crystal-coil transition may exceed the 
size of a small system. Figure~\ref{fig:fcs_S50T25} demonstrates
that, for a polymer with bending stiffness $\epsilon_{\rm b}=5.0$ at temperature
$T=2.5$, only chains with length $N \geq  1,000$ will show clear
signature of a
discontinuous structural transition. When the chain length $N$ is
equal to $400$ or $500$, at the 
crystal-coil transition force (which corresponds to the
maximum of the extension susceptibility $\chi_f$),
the two peaks of the distribution of $n_{\rm c}$ have the trend of
merging into one peak. For these later small systems, the crystal-coil
transition resembles a continuous phase-transition. 

According to the phase-diagram
Fig.~\ref{fig:temperature-delta-phase-diagram}, at fixed temperature
$T$, there exists a threshold bending stiffness value $\epsilon_{\rm
b}^{*}(T)$ at which the force-induced collapse transition
changes from being second-order to being first-order. To determine
the value of $\epsilon_{\rm b}^{*}$ precisely, however, is not easy.
When the bending stiffness of the polymer is just slightly above
$\epsilon_{\rm b}^{*}$, although an infinite chain will have a
jump in the order parameters $n_{\rm c}$ and $n_{\rm b}$
at the collapse
transition point, for an finite chain such jumps in $n_{\rm c}$
and $n_{\rm b}$ will be smeared out by large fluctuations of 
$N_{\rm c}$ and $N_{\rm b}$ at the
transition point. For a polymer of length $N$ less than or
comparable to the correlation length of the system at the
first-order collapse transition point, the elongation of the polymer
may still follow the scaling rule Eq.~(\ref{eq:X-N-scaling}), i.e.,
resembles a continuous structural transition. At $T=1.2$, the
critical exponent $\mu_{\rm cr}$ (defined in the previous
subsection) as a function of $\epsilon_{\rm b}$ is shown
in Fig.~\ref{fig:Scaling_Stiff_Force} for polymers of 
length up to
$N=1,000$. We find that $\mu_{\rm cr}$ first decreases with 
$\epsilon_{\rm  b}$; but for $\epsilon_{\rm b} > 1.0$ it increases with
$\epsilon_{\rm b}$. We interpret this change of trend as follows: At
a continuous collapse transition, the stiffer the chain is, the more
compact it prefers and therefore the smaller the critical exponent
$\mu$ is; on the other hand, at a discontinuous collapse transition,
the stiffer the chain is, the more extended the average elongation
of the polymer is and the larger the value of $\mu_{\rm cr}$ is. Based on
this argument, we estimate the value of $\epsilon_{\rm b}^{*}$ as
the point where the scaling exponent $\mu_{\rm cr}$ reaches the minimal
value.


\section{Summary and discussion}
\label{sec:conclusion}

The qualitative properties of temperature- and force-induced
collapse transitions of a 2D self-attractive semiflexible lattice
polymer were investigated by Monte Carlo simulations. The system has
three possible phases: the crystal phase, the disordered globular
phase, and the random coil phase. The crystal-globule and the
crystal-coil transitions are both first-order phase-transitions in
the thermodynamic limit; while the globule-coil transition is a
second-order phase transition. The disordered globular phase is
absent for polymers of considerable stiffness. These simulation
results are consistent with earlier theoretical
\cite{Doniach-etal-1996} and simulation
\cite{Bastolla-Grassberger-1997,Doye-etal-1998} studies on
temperature-induced collapse transitions of 3D lattice polymers.
They also confirm and extend the earlier simulation
\cite{Grassberger-Hsu-2002} and analytical
\cite{Marenduzzo-etal-2004,Zhou-etal-2006} studies on force-induced
collapse transitions of 2D lattice polymers.

The simulation results also suggested that, at the continuous
force-induced globule-coil transition, the extension of the polymer
scales with polymer length according to $\langle X_N \rangle =
a_0 N^{\frac{\mu_{\rm cr}}{2}}$, with both $a_0$ and $\mu_{\rm cr}$ being
independent of chain length. Such a critical behavior is similar to
the well-known critical scaling $R_g \propto a_1 N^{\nu_{\rm cr}}$ at the
temperature-induced globule-coil transition [see Eq.~(\ref{EQ:power-law})],
but it is yet to be confirmed by analytical calculations.
According to the simulation results, the bending stiffness does not affect the
critical exponent $\nu_{\rm cr}$.  However,
the scaling exponent $\mu_{\rm cr}$  for the
force-induced globule-coil transition changes with the bending
stiffness $\epsilon_{\rm b}$ (see
Fig.~\ref{fig:Scaling_Stiff_Force}). We suggest such a qualitative
difference should be amenable to experimental verifications
\cite{Maier-Raedler-1999,Lin-etal-2007}. 
A possible reason for this qualitative
difference were also given in this paper.

For 2D polymers, we have shown that the temperature- and
force-induced collapse transitions have the same order. This may not
be the case for 3D polymers. For 3D flexible polymers, the
force-induced globule-coil transition is a first-order
phase-transition \cite{Grassberger-Hsu-2002}; the order of the
temperature-induced collapse transition, on the other hand,
 is usually regarded as second-order in
many references (but Binder and co-authors \cite{Rampf-etal-2005}
argued that it is actually also a first-order phase-transition).


\section*{Acknowledgment}

We are grateful to Luru Dai and Zhihui Wang for their helps on
simulation techniques and programming. The computer simulation was
performed at the PC clusters of the State Key Laboratory of
Scientific and Engineering Computing, Beijing.



\end{document}